\documentclass[reprint,superscriptaddress,amsmath,amssymb,aps]{revtex4-2}
\usepackage[utf8]{inputenc}
\usepackage{amsmath,amsthm,amssymb,amsfonts,braket}
\usepackage{placeins}[verbose]
\usepackage{tabularx, longtable, multirow}
\usepackage{graphicx}
\usepackage{dcolumn}
\usepackage{bm,dsfont}
\usepackage[english]{babel}
\usepackage{comment}
\usepackage[dvipsnames]{xcolor}
\usepackage{siunitx}

\def\bk{{\mathbf{k}}}

\def\bp{{\mathbf{p}}} 
 
\def\br{{\mathbf{r}}}
\def\bxi{{\boldsymbol \xi}}

\begin{document}
\title{Neutron State Entanglement with Overlapping Paths}

\author{S. J. Kuhn$^*$}
 \affiliation{Center for Exploration of Energy and Matter, Indiana University, Bloomington, 47408, USA}
\thanks{These authors contributed equally to this work.}
\author{S. McKay$^*$}
 \affiliation{Center for Exploration of Energy and Matter, Indiana University, Bloomington, 47408, USA} 
\author{J. Shen}
 \affiliation{Center for Exploration of Energy and Matter, Indiana University, Bloomington, 47408, USA}
\author{N. Geerits}
 \affiliation{Atominstitut, TU Wien, Stadionallee 2, 1020 Vienna, Austria}
\author{R. M. Dalgliesh}
 \affiliation{ISIS, Rutherford Appleton Laboratory, Chilton, Oxfordshire, OX11 0QX, UK}
\author{E. Dees}
 \affiliation{Center for Exploration of Energy and Matter, Indiana University, Bloomington, 47408, USA}
\author{A. A. M. Irfan}
\affiliation{Department of Physics, Indiana University, Bloomington IN 47405, USA}
\author{F. Li}
\affiliation{Neutron Sciences Directorate, Oak Ridge National Laboratory, Oak Ridge, TN, 37830, USA}
\author{S. Lu}
\affiliation{Department of Physics, Indiana University, Bloomington IN 47405, USA}
\author{V. Vangelista}
 \affiliation{Center for Exploration of Energy and Matter, Indiana University, Bloomington, 47408, USA}
\author{D. V. Baxter}
 \affiliation{Center for Exploration of Energy and Matter, Indiana University, Bloomington, 47408, USA}
\affiliation{Quantum Science and Engineering Center, Indiana University, Bloomington, IN 47408, USA}
\author{G. Ortiz}
\affiliation{Department of Physics, Indiana University, Bloomington IN 47405, USA}
\affiliation{Quantum Science and Engineering Center, Indiana University, Bloomington, IN 47408, USA}
\author{S. R. Parnell}
 \affiliation{Faculty of Applied Sciences, Delft University of Technology, Mekelweg 15, 2629 JB Delft, The Netherlands}
\author{W. M. Snow}
 \affiliation{Center for Exploration of Energy and Matter, Indiana University, Bloomington, 47408, USA}
\affiliation{Quantum Science and Engineering Center, Indiana University, Bloomington, IN 47408, USA}
\author{R. Pynn}
\affiliation{Center for Exploration of Energy and Matter, Indiana University, Bloomington, 47408, USA}
\affiliation{Neutron Sciences Directorate, Oak Ridge National Laboratory, Oak Ridge, TN, 37830, USA}
\affiliation{Quantum Science and Engineering Center, Indiana University, Bloomington, IN 47408, USA}

\date{\today}

\begin{abstract}
The development of direct probes of entanglement is integral to the rapidly expanding field of complex quantum materials. 
Here we test the robustness of entangled neutrons as a quantum probe by measuring the Clauser-Horne-Shimony-Holt contextuality witness while varying the beam properties.
Specifically, we prove that the entanglement of the spin and path subsystems of individual neutrons prepared in two different experiments using two different apparatuses persists even after varying the entanglement length, coherence length, and neutron energy difference of the paths. The two independent apparatuses acting as entangler-disentangler pairs are static-field magnetic Wollaston prisms and resonance-field radio frequency flippers.
Our results show that the spatial and energy properties of the neutron beam may be significantly altered without reducing the contextuality witness value below the Tsirelson bound, meaning that maximum entanglement is preserved. We also show that two paths may be considered distinguishable even when separated by less than the neutron coherence length. This work is the key step in the realization of the new modular, robust technique of entangled neutron scattering.
\end{abstract}

\maketitle

\section{Introduction}

Advancing the frontiers of science often requires the creation of new physical methods to uncover the underlying microscopic mechanisms that give rise to exotic macroscopic phenomena. A myriad of scattering techniques, based on photon, electron, X-ray, or neutron probes, are currently being used with great success to discover and characterize fundamental properties of complex materials. These probe techniques base their success on the control and manipulation of two of the defining traits of quantum mechanics, namely, discreteness of elementary physical properties and interference phenomena, allowing inference of certain spacetime correlations of the target sample.
However, direct measurement of quantum entanglement within complex materials remains elusive. Often this entanglement is thought to be at the root of the underlying microscopic mechanisms that give rise to remarkable phenomena such as emergent chirality in spin liquids, topological quantum order, strange metallic behavior, and unconventional superconductivity. A new type of probe that exploits entanglement, a uniquely quantum resource, may help directly reveal some of these phenomena. 

In a previous experimental paper, we introduced a fundamentally new quantum probe, a beam of mode-entangled (i.e. intraparticle-entangled) neutrons \cite{shen2019}. In that experiment we proved neutron multimode-entanglement by demonstrating a violation of Bell-type inequalities for both bipartite (spin and path) and tripartite (spin, path, and energy) distinguishable subsystems. Path refers to the neutron spatial trajectory along the instrument. Those experiments, performed at the ISIS muon and neutron facility, used one type of neutron subsystem-entangler, a pair of radio-frequency (RF) flippers.
The RF flippers refract the neutron's up and down spin states into spatially separated, parallel path states, effectively splitting the single neutron into two separate two-state subsystems. The spatial separation between the paths is defined as the entanglement length $\xi$ (also called the spin echo length). 

A recent theoretical investigation has shown that a spin-path entangled neutron probe has unique and complex scattering signatures from interactions with an entangled target state, at various length scales, in toy models of magnetic materials \cite{Irfan2020}. A typical length scale of entanglement in quantum materials is on the order of tens of nanometers or smaller, which is much less than the neutron entanglement length of 1500 nm used in the previous experiment \cite{shen2019}. That previous work also paid no attention to the coherence properties of the neutrons, nor to the explicit proof of the distinguishability of the two path states. The present work addresses the relationship between the neutrons' entanglement and coherence properties by exploring a wide range of relevant length scales of the entangled neutron beam.
Our argument of universal applicability hinges on the degree of tunability of our entangled neutron probe.

Our model for neutron coherence corresponds to that recently articulated in detail in \cite{Majkrzak2019}: the neutron beam consists of individual neutrons, uncorrelated with one another, each having a transverse spatial extent that defines the area of a sample with which the neutron interacts coherently; this spatial extent is the \textit{transverse intrinsic coherence length} $\Delta_t$. This intrinsic coherence length is also often taken to be the transverse size of the individual neutron wavepacket, although this need not necessarily be the case as linear wavepackets spread during propagation, while the transverse intrinsic coherence length remains constant \cite{klein-1983}. The wavepacket size is presumably determined by the way in which each neutron is produced, both by nuclear reactions at the source and by scattering within a neutron moderator or from a crystal monochromator.

\begin{figure}[t]
\includegraphics[width=3.4in]{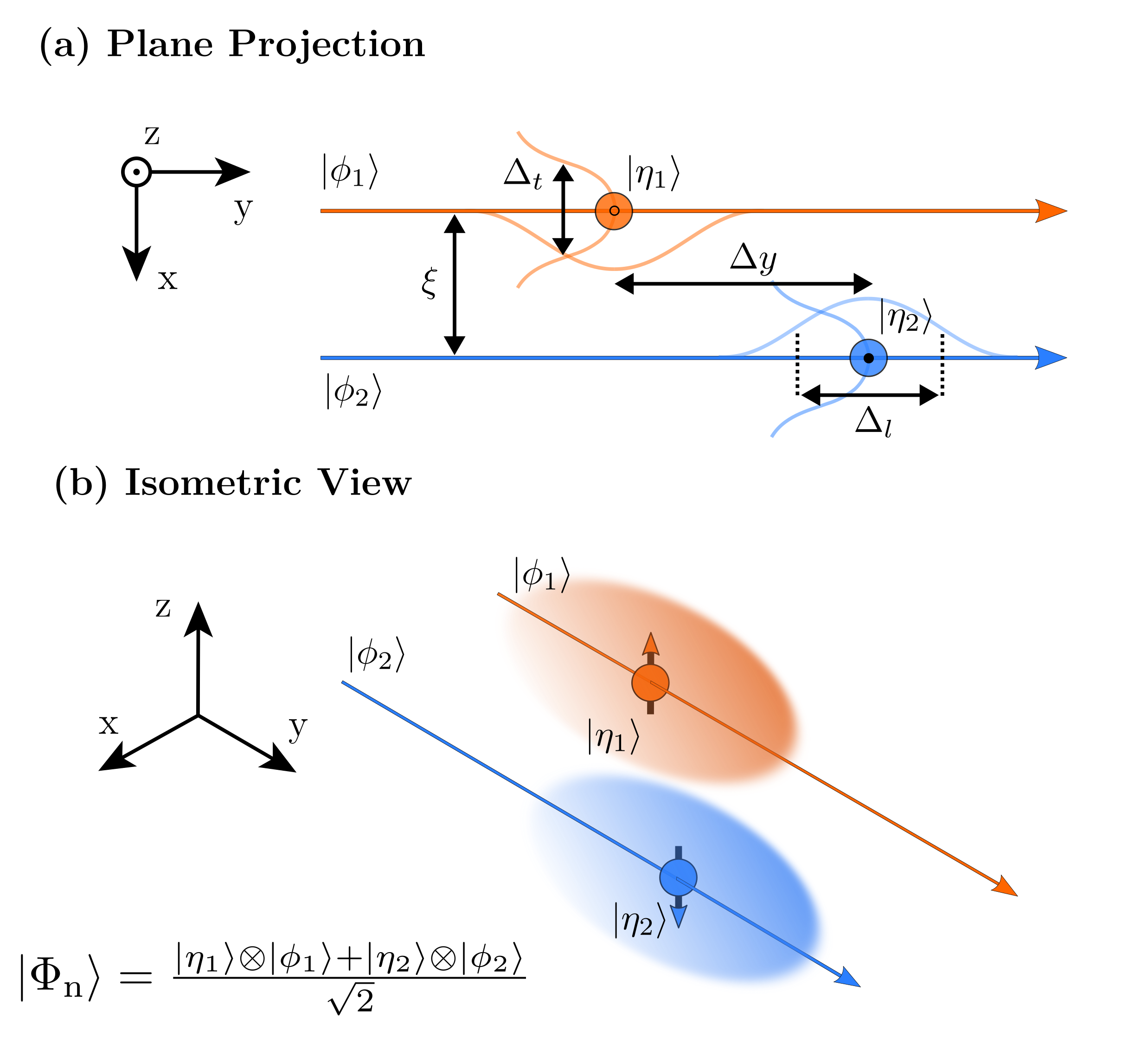}
\caption{\label{fig:wavep} Definitions of intrinsic coherence volume of a single neutron, where $\Delta_t$ is the transverse coherence length, $\Delta_l$ the longitudinal coherence length, $\Delta y$ the longitudinal peak separation of the wavepacket branches, and $\xi$ the entanglement length. The kets $\ket{\eta_i}$ and $\ket{\phi_i}$ for $i = 1,2$ denote the spin and path states, respectively. The total neutron state $\ket{\Phi_{\rm n}}$ is mode-entangled [see Eqn. \eqref{Eqn:generalstate}]. The thick orange and blue lines are the ``classical" trajectories of the neutron, although classically a neutron cannot simultaneously take both routes. The orange and blue shaded regions in (b) represent the intrinsic coherence volume. }
\end{figure}

Expanding on the concept of coherence length, we note that all neutron sources produce neutrons that are mutually incoherent; the coherence volume is typically defined by the slit geometry and configuration of the instrument. In our present experiments, we are primarily interested in coherence in directions perpendicular to the neutron's propagation.
We will refer to the transverse coherence length defined by a slit as the \textit{transverse beam coherence length} $\beta_t = \ell / (k_n a)$ for a neutron wavevector of magnitude $k_n$ defined by a slit of width $a$, with distance $\ell$ between the slit and the point of interest on the axis of propagation \cite{keller1997}. This finite length, which is a beam property rather than an individual neutron property, gives rise to an observed broadening of the interference pattern produced by scattered neutrons that is often referred to as the result of the finite resolution of the neutron instrumentation. In our experiments, we only had direct experimental control of $\beta_t$ and not $\Delta_t$. 
However, recent experiments have suggested that the transverse intrinsic coherence length for a single neutron is much larger than the beam coherence length ($\Delta_t \gg \beta_t$) \cite{Majkrzak2019,treimer2006}; by definition, $\Delta_t \geq \beta_t$.
In most neutron scattering experiments, typical values for $\beta_t$ range from 50 nm to 500 nm while the entanglement length $\xi$ may be tuned between a few tens of nanometers and several microns \cite{li2014}.

In addition to the transverse coherence length one can also define several other coherence lengths for the assembly of neutrons in a beam, in the traditional manner used for light optics. For example, the longitudinal beam coherence length is defined by the degree of beam monochromatization and is traditionally written as $\beta_l = \lambda_n^2 / \Delta \lambda_n$, where $\lambda_n$ is the neutron wavelength and $\Delta \lambda_n$ its uncertainty, which is approximately determined by the pulse width at pulsed neutron sources (\SI{300}{\micro\second} at ISIS) or by the monochromator at a continuous neutron source  \cite{keller1997}.
As in the transverse case, we currently can only experimentally measure and control $\beta_l$ and not the longitudinal intrinsic coherence length $\Delta_l$.
Furthermore, we must also consider the longitudinal overlap of the path states in those experiments where entangled neutrons are produced using RF flippers; the slight velocity difference between the two path states leads to a longitudinal separation $\Delta y$ of the two states (see Fig. \ref{fig:wavep}). We seek to determine whether the subsystem distinguishability assumption depends on this separation as well.

On qualitative grounds, when the entanglement length is much longer than the transverse intrinsic coherence length ($\xi \gg \Delta_t$), one can consider the two path states as distinguishable;  traditional neutron interferometry is always within this regime. As the entanglement length is reduced below the neutron coherence length ($\xi \leq \Delta_t$), will the assumption of distinguishability of the path subsystem still hold? This key question presents a theoretical hurdle as well as experimental challenges. Theoretically, the neutron wavepacket's spatial subsystem must be written using an uncountable basis for the path subsystem, but the Clauser, Horne, Shimony, and Holt (CHSH) contextuality inequality requires distinguishable, discrete basis subsystems for both spin and path. Experimentally, an entangled neutron probe would be much easier to implement if it is not sensitive to small changes in the neutron beam coherence.

To prove the robustness of the entanglement in the neutron probe, it is important on foundational grounds to understand the breakdown of the subsystem distinguishability assumption --- in which the two paths are taken as distinct quantum modes \cite{lu2019operator} ---  as one varies the transverse intrinsic coherence length $\Delta_t$ and the entanglement length $\xi$, as shown in Fig. \ref{fig:wavep}. The experiments described in this paper address the effect of beam coherence on the degree of entanglement as determined by the value of the CHSH contextuality witness $S$ defined in Sec. \ref{CHSHwitness}.
Two complementary experiments were performed to measure the contextuality witness $S$ for neutrons with the spin and path subsystems entangled: magnetic Wollaston prisms (MWP) were used at the High Flux Isotope Reactor (HFIR) at Oak Ridge National Laboratory, and Radio Frequency (RF) flippers were employed at the ISIS pulsed-neutron source. In both cases, the relative phases between the two states of both the spin and path subsystems were independently manipulated to determine $S$ as defined in Eqn. \eqref{Eqn:CHSH Witness}. Our experiments focused on the case where the two paths significantly overlap in the transverse direction ($\xi < \beta_t \leq \Delta_t$). In our original work \cite{shen2019}, we were in the regime of $\xi >\beta_t$, which left open the possibility that $\xi > \Delta_t$. Therefore, these new measurements determine if spatial separation is required for path state distinguishability and thus neutron entanglement.

Our experimental results indicate that the entanglement between spin and path subsystems is quite robust across different entangling devices, with overlapping path states and varying neutron velocity and wavelength having no effect on the degree of entanglement. Therefore, the distinguishability assumption for the paths holds over a wide range of experimental conditions. It is precisely this flexibility and wide tunability of the entanglement length $\xi$ that make entangled-neutron scattering techniques potentially attractive for studies over a wide range of length scales.

\section{Methods}

\begin{figure*}[t]
\includegraphics[width=7in]{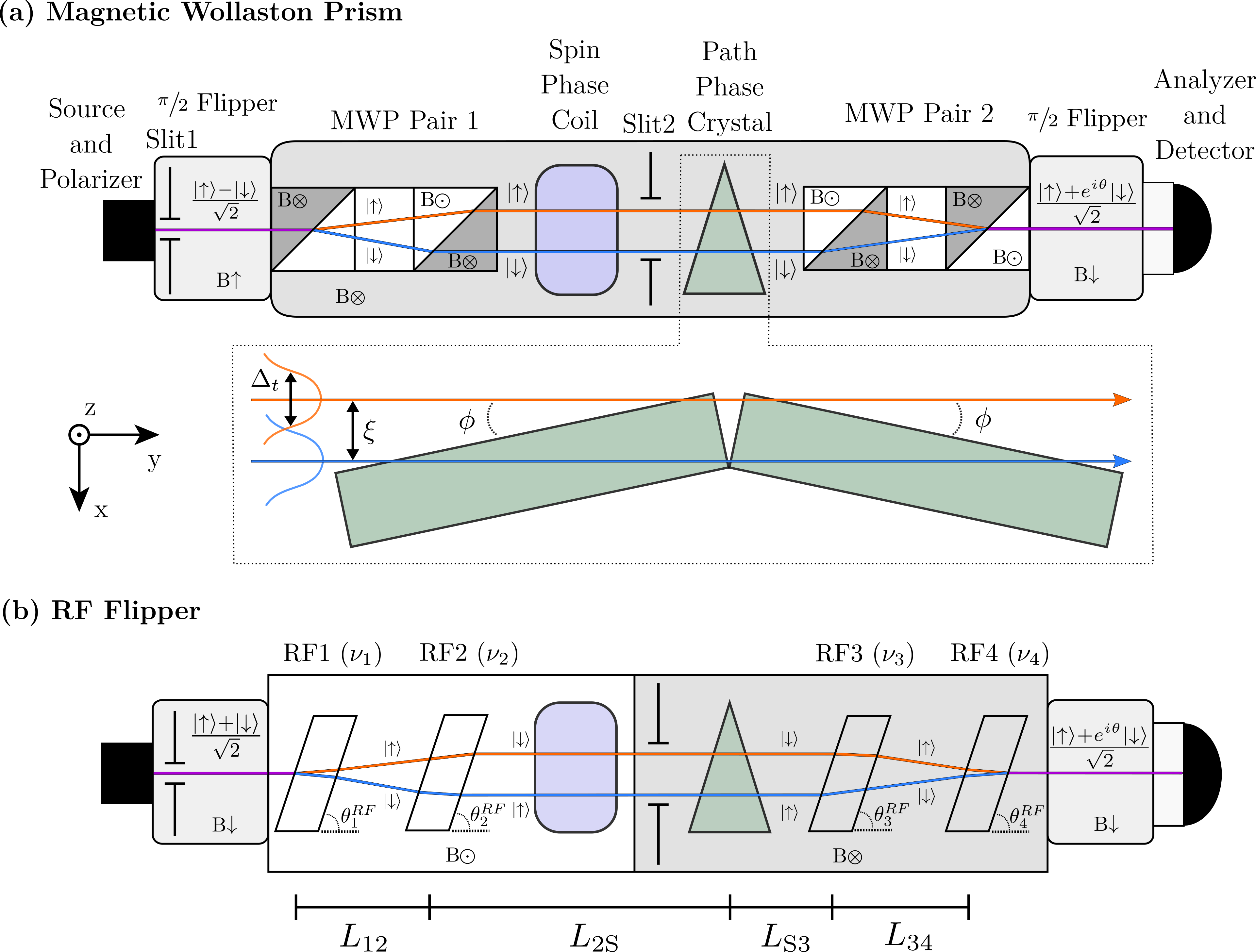}
\caption{\label{fig:instrument}Schematic of the instrument setup utilizing (a) the magnetic Wollaston prism (MWP) pairs and (b) RF flippers. In both cases, the $z$-polarized neutron beam travels in the $y$ direction from the left side of the diagram (the source and polarizer) to the right side (the polarization analyzer and detector).
The beam is always within a magnetic field.
The $\pi / 2$ flipper orients the neutron polarization along the $x$ direction, corresponding to a superposition of the up and down states in the $z$-basis. Slit1 constrains the transverse beam size.
(a) In the MWP setup, the neutrons enter all MWPs non-adiabatically. Each MWP contains a 45$^\circ$ border where the field changes from $-z$ (dark grey) to $+z$ (white), causing the up and down spin states to refract in opposite directions, but they remain in the $x$-$y$ plane. A second 45$^\circ$ border refracts the two states so they are traveling parallel with one another, but with paths separated by the entanglement length $\xi$. Both MWP pairs are separated by a rectangular field. The beam passes through the spin-phase coil that produces a magnetic field along $z$, which tunes the spin phase. The beam dimensions are further constrained by Slit2, which is much wider than the entanglement length. Next, the neutron passes through two quartz blocks placed as shown in the insert, whose orientation results in a path phase difference between the two separated paths. The beam is then spatially recombined by the second MWP pair, before having a single spin state chosen by the polarization analyzer and the intensity measured by the detector. 
(b) In the RF flipper setup, RF flippers replace the MWPs but otherwise the setup remains essentially the same. Each RF flipper is tuned to a frequency $\nu$ and is placed at an angle $\theta^{RF}$ relative to the $y$-axis. The static field of the RF flipper is in the same direction as the guide field, which is reversed before Slit2 in conventional mode so as to give no net field integral. The definitions of the lengths between the RF flipper centers and the center of the quartz blocks are shown.}
\end{figure*}

\begin{figure*}[t]
\includegraphics[width=5.5in]{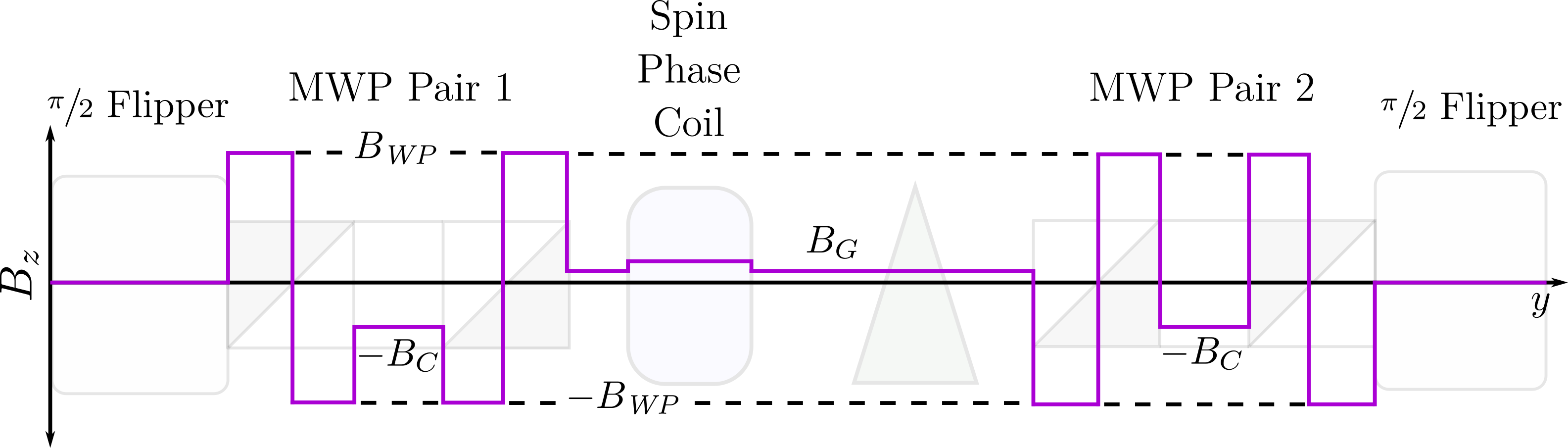}
\caption{\label{fig:MWPStaticField}The magnetic field in the vertical $z$ direction ($B_z$) along the magnetic Wollaston prism (MWP) setup. Field strength is not to scale. The instrument components are shown in the background to indicate the position along the beamline. Inside the $\pi / 2$ flipper, the field is in the $x$ direction. The magnetic field in each triangle of the MWP takes a value of $\pm B_{WP}$. In the central field of each MWP, a field of $-B_C$ is applied. An external guide field $B_G$ is applied between the MWP pairs, with a small additional field optionally provided by the spin-phase coil. All changes in the field direction are non-adiabatic.}
\end{figure*}

In both the HFIR and ISIS experiments, the relative spin phase $\alpha$ and path phase $\chi$ between the two states of the two neutron subsystems were tuned by applying a small additional $z$-directed magnetic field or by passing neutrons through one or more inclined quartz blocks, respectively. Single-crystal quartz blocks were chosen for their relatively large coherent neutron scattering length and small neutron absorption. 
In the MWP experiment, two quartz blocks were mounted on separate 360$^\circ$ rotation stages, while in the RF flipper experiment, quartz blocks were mounted on a table at a series of predetermined angles. 

The equation for relative path phase for both experiments is 
\begin{equation}
    \label{Eqn:pathphase}
    \chi = m \lambda_n \xi \rho (\cot(\phi)+\tan(\phi)) = \frac{2m \lambda_n \xi \rho}{\sin(2\phi)} ,
\end{equation}
where $m$ is the number of quartz blocks (usually 2 or 4), $\rho$ the  scattering length density of quartz \cite{NIST_SLDcalculator}, and $\phi$ the angle the blocks make with the $y$-axis. The blocks were placed as shown schematically in the insert of Fig. \ref{fig:instrument}(a), with blocks rotated by $\pm \phi$ to form a tent-like configuration in order to reduce the error in path phase due to a divergent beam.
With an even number of blocks, the dependence of the relative path phase on the angular divergence of the beam $\delta \phi$ is $\chi(\phi + \delta \phi) = \chi(\phi) + \mathcal{O}(\delta \phi^2)$.

The relative spin phase due to the spin-phase coil is found via the equation
\begin{equation}
    \label{Eqn:spinphase}
    \alpha = C_{\alpha} \lambda_n B_{\alpha} d
\end{equation}
where $\lambda_n$ is the wavelength of the neutron, $B_{\alpha}$ the magnetic field supplied by the spin-phase coil, $d$ is the distance traveled by the neutron in the spin-phase coil's field, and $C_{\alpha} = 4.632 \times 10^{14}$ T$^{-1}$ m$^{-2}$.
This relative phase is due to the rotation of the spin about the applied magnetic field; classically, this rotation is understood as Larmor precession. 

\subsection{Magnetic Wollaston prism experiment}

A constant neutron wavelength measurement was performed on the CG-4b beamline at HFIR using the apparatus sketched in Fig. \ref{fig:instrument}(a) and described in detail in the figure caption. The experiment used a monochromatic beam of neutrons of wavelength $\lambda_n=5.4$ {\AA} reflected from a silicon crystal monochromator and polarized in the vertical ($z$) direction by a s-bender. A low-efficiency beam monitor was placed directly before the first beam profile-defining slit (denoted ``Slit1'' in Fig. \ref{fig:instrument}), which was selected from an array of slits, each 10 mm tall, with widths 0.5, 2, and 4 mm.  

A pair of MWPs act as a spin and path entangler by transversally separating the two neutron spin states into two outgoing path states \cite{li2014}. Within each MWP, superconducting-wire triangular coils produce static fields in the $+z$ or $-z$ directions. Neighboring triangular regions are separated by a high-temperature superconducting (HTS) film inclined at 45$^{\circ}$ to the neutron trajectory, as shown by Fig. \ref{fig:instrument}(a). Two MWPs, separated by a rectangular-shaped static magnetic field, constitute a MWP pair. The fields in the rectangular regions are tuned independently to obtain a net magnetic field path-integral of zero between the two ends of the apparatus. 
When entering the MWP, neutrons are in a superposition of the up and down spin states, defined along the $z$-axis. The two spin states refract in opposite directions when the field is reversed abruptly at the inclined field boundary within an MWP. The two paths are made parallel again by the second MWP. The transverse separation of the two paths $\xi$ is determined by the strength of the fields in the MWPs as well as by the separation of the MWPs, as shown in the following equation
\begin{equation}
\label{Eqn:MWP_SEL}
\xi = C_{\xi}\lambda_n^2 B_{WP} L \cot{\theta_f} ,
\end{equation}
where $\xi$ is the entanglement length, $\lambda_n$ the wavelength of the neutron, $B_{WP}$ the magnetic field in the MWP,  $L$ the distance between MWPs centers (0.21 m), $\theta_f$ the angle of the film to the beam (45$^{\circ}$), and $C_{\xi}=1.474\times 10^{14}$ T$^{-1} \cdot$m$^{-2}$ \cite{li2014}.

The second MWP pair has static field directions reversed with respect to the first MWP pair, causing the spatially separated states to interfere before the $x$-component of the neutron polarization is selected at the exit of the final MWP and passed to a supermirror polarization analyzer and a $^3$He neutron detector. In the absence of the quartz crystal shown in Fig. \ref{fig:instrument}, the beam polarization is brought to the $x$ direction (i.e. spin echo is achieved) at the exit of the final MWP by adjusting the strengths of the rectangular shaped field regions within the MWPs. Together with Slit1, a 2-mm-wide, 10-mm-tall slit (Slit2) between the spin-phase coil and the path-phase crystal defines the beam divergence. 

\subsection{RF flipper experiment}
This experiment was conducted using the Larmor instrument located at the second target station of the ISIS neutron and muon source, part of the Rutherford Appleton Laboratory (RAL) in the UK. 
The primary difference between this experiment and the one using MWPs is that the spin and path subsystems are entangled and disentangled by two pairs of RF flippers whose angles relative to the beam $\theta^{RF}$ and frequencies $\nu$ are adjustable, as shown in Tab. \ref{tab:RFParams}. Like in the MWP experiment, each neutron is in a superposition of the up and down spin states when it enters the first RF flipper (RF1).

The RF flippers produce static magnetic fields in the $\pm z$ direction that satisfy the resonance condition for the chosen RF frequency. Unlike in the MWP experiment, in which the neutrons maintain a constant energy, during the RF $\pi$-flip each neutron spin state experiences a small change in its total energy. The magnitude of the RF field is varied during each neutron pulse to ensure that a $\pi$-flip is achieved for all neutron wavelengths between 3.5 and 7.5 {\AA} \cite{1/tThing}. Similar to the MWP, when the boundary of the static field of the flipper is at angle $\theta^{RF}$ to the neutron beam, the two paths will be separated along the $x$ direction by a distance $\xi$. The third and fourth RF flippers disentangle the spin and path subsystems before an analyzer and detector similar to those used in the MWP setup.

\begin{figure}[t]
\includegraphics[width=3.4in]{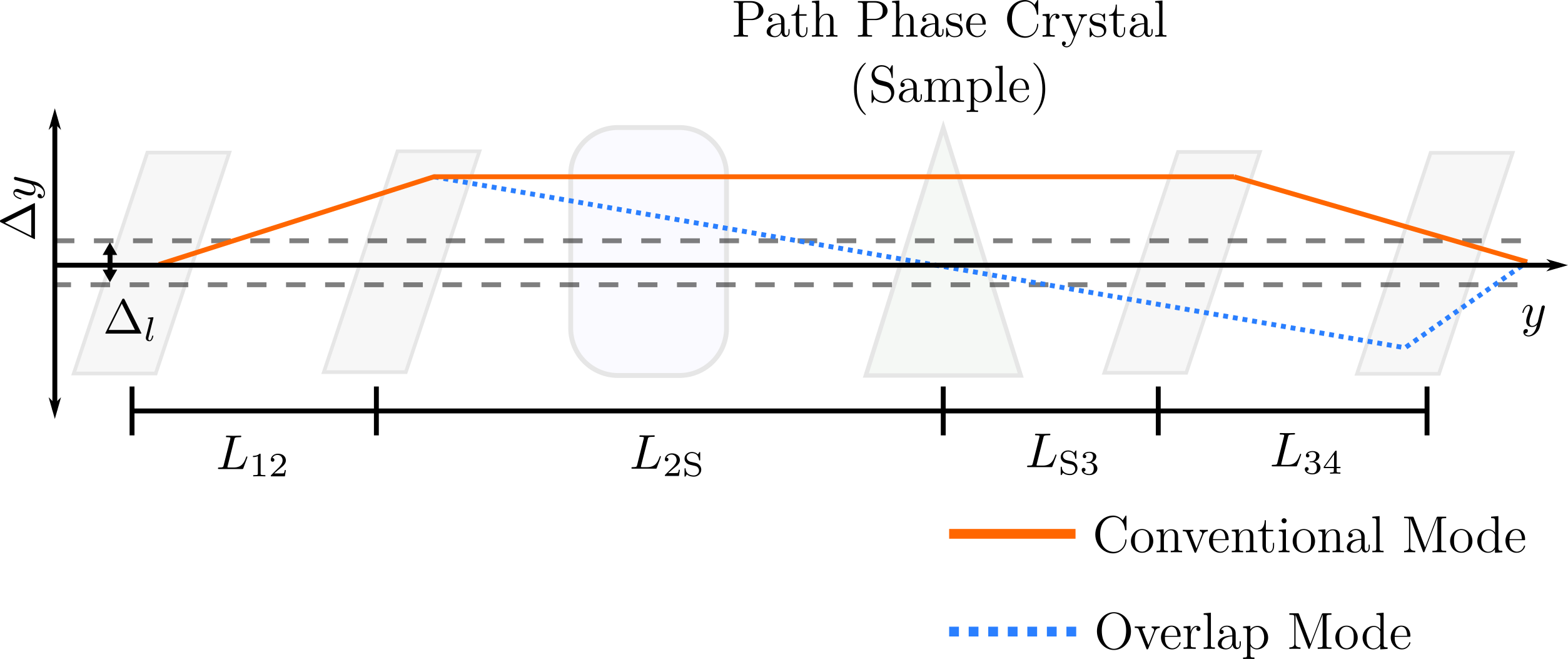}
\caption{\label{fig:overlap}The longitudinal separation $\Delta y$ of the leading and lagging wavepackets as they proceed along the beamline. The energy change caused by the RF flippers leads to a velocity difference between the states and hence to a longitudinal spatial separation that increases between RF1 and RF2. In conventional mode, RF1 and RF2 have the same frequency, leading to a constant separation between RF2 and RF3. In overlap mode, RF1 has a lower frequency than RF2, causing the separation to decrease between RF2 and RF3. The angles of the flippers with respect to the neutron beam are adjusted to keep the entanglement length constant, even when the frequencies of the flippers are changed. The longitudinal coherence length $\Delta_l$ of the neutron is indicated by the dashed lines.}
\end{figure}

Unlike the MWPs, the RF flippers slightly change the total energy for each neutron spin state, adding energy to one state and removing from the other during the spin flip process. Accordingly, the neutron states have different kinetic energies when they leave RF1, resulting in an increasing longitudinal separation between the two spin components between the first two RF flippers, as shown by the conventional mode line in Fig. \ref{fig:overlap}. In this conventional mode, the second RF flipper has the same frequency as the first, undoing the kinetic energy change introduced by the first flipper and resulting in a constant longitudinal separation of the two states when they interact with the quartz blocks. 
In our experiment, this separation can be appreciable, reaching about 400 nm for 4 {\AA} neutrons with an RF frequency of 500 kHz and a distance of roughly 1 m between the first two RF flippers. This separation is substantially larger than the expected longitudinal coherence length of the neutron beam given by $\beta_l$, where $\beta_l \approx 20$ nm for the same neutrons.  However, the RF flipper frequencies can be adjusted such that there is a velocity difference between the two wavepacket branches between RF2 and RF3, allowing them to overlap as they pass through the quartz crystals. Because the RF flipper frequencies are directly tied to the static magnetic fields, changing the frequency also changes the refraction angle for each spin state. The inclination of RF2 thus has to be modified to ensure that the paths are parallel between RF2 and RF3. The guide field direction is not flipped between RF2 and RF3 in overlap mode, but rather the same direction throughout.

Appropriate changes are made to the angles and frequencies of RF3 and RF4 to ensure that the two wavepacket branches interfere at the detector; the solution to the following set of equations is the focusing condition:
\begin{subequations}
\begin{align}
\nu_2 &= \frac{L_{12} +L_{2{\rm S}}}{L_{2{\rm S}}}\nu_1 \\
\nu_3 &= \frac{L_{{\rm S}3} +L_{34}}{L_{34}}(\nu_2-\nu_1) \\
\nu_4 &= \nu_3 - \nu_2 + \nu_1
\end{align}
\end{subequations}
As shown in Fig. \ref{fig:overlap}, each $L_{jk}$ is one of the various distances between the RF flippers (1, 2, 3, 4) and the sample position (S). Here $\nu_1$ was selected to give the desired value of entanglement length. The RF flipper angles $\theta^{RF}_i$ (with $i=1,2,3,4$) were selected to maintain a constant entanglement length. In this \textit{overlap mode} configuration, the wavepacket branches are within the longitudinal coherence length of one another within the path-phase crystals.
All relevant RF flipper parameters used to ensure the focusing condition are in Tab. \ref{tab:RFParams}.

\begin{table}[ht]
\centering
\caption{\label{tab:RFParams}RF flipper parameters.}
\begin{ruledtabular}
\newcolumntype{R}{>{\centering\arraybackslash}X}
\begin{tabularx}{0.3\linewidth}{R|R|R}
    Parameter & Conventional Mode & Overlap Mode \\
\hline
{\begin{tabularx}{\linewidth}{R|R} 
    \multirow{4}{*}{(kHz)} & $\nu_1$  \\
    & $\nu_2$ \\
    & $\nu_3$ \\
    & $\nu_4$ \\
\end{tabularx}}
& {\begin{tabularx}{\linewidth}{R}
    500 \\ 500 \\ 500 \\ 500 \\
\end{tabularx}}
& {\begin{tabularx}{\linewidth}{R}
    600 \\ 902 \\ 575 \\ 273 \\
\end{tabularx}} \\
\hline
{\begin{tabularx}{\linewidth}{R|R} 
    \rule{0pt}{2.5ex} 
    \multirow{4}{*}{(deg)} & $\theta^{RF}_1$ \\
    & $\theta^{RF}_2$ \\
    & $\theta^{RF}_3$ \\
    & $\theta^{RF}_4$ \\
\end{tabularx}}
& {\begin{tabularx}{\linewidth}{R}
    70 \\ 70 \\ 70 
    \\ 70 
    \\
\end{tabularx}}
& {\begin{tabularx}{\linewidth}{R}
    80.0 \\ 83.3 \\ 124.4 
    \\ 113.3 
    \\
\end{tabularx}} \\
\hline
{\begin{tabularx}{\linewidth}{R|R}
    \rule{0pt}{2.5ex} 
    \multirow{4}{*}{(m)} & $L_{12}$ \\
    & $L_{2\mathrm{S}}$ \\
    & $L_{\mathrm{S}3}$ \\
    & $L_{34}$ \\
\end{tabularx}}
& {\begin{tabularx}{\linewidth}{R}
     1.20 \\ 2.383 \\ 1.065 \\ 1.18 \\
\end{tabularx}}
& {\begin{tabularx}{\linewidth}{R}
    1.20 \\ 2.383 \\ 1.065 \\ 1.18 \\
\end{tabularx}} \\
\end{tabularx}
\end{ruledtabular}
\end{table}

\section{Results}

\begin{table*}[t]
\centering
\caption{\label{tab:WitValues}Witness values with beam coherence and entanglement lengths. Statistical uncertainties for the measured coherence lengths and the entanglement length were 5$\%$ and 1$\%$, respectively. }
\begin{ruledtabular}
\newcolumntype{R}{>{\centering\arraybackslash}X}
\begin{tabularx}{0.15\linewidth}{R|R|R|R|R|R}
Experiment  &  Coherence Length $\beta_t$  (nm) & Entanglement Length $\xi$ (nm) & Polarization & Experimental Witness Value & Maximum Entangled Witness Value ($2\sqrt{2} \times$Pol) \\
\hline
RF Conv. \cite{shen2019} & 100 & 1600 & 0.78 $\pm$ 0.02 & 2.16 $\pm$ 0.02 & 2.20 $\pm$ 0.06 \\
MWP 0.5 mm  & 550 & 600 & 0.86 $\pm$ 0.03 & 2.50 $\pm$ 0.01 & 2.43 $\pm$ 0.08 \\
MWP 2 mm  & 140 & 600 & 0.89 $\pm$ 0.02 & 2.50 $\pm$ 0.01 & 2.51 $\pm$ 0.06\\
MWP  4 mm  & 70 & 600 & 0.88 $\pm$ 0.02 & 2.50 $\pm$ 0.01 & 2.48 $\pm$ 0.06 \\
RF Conv.   & 350 & 85 & 0.85 $\pm$ 0.02 & 2.42 $\pm$ 0.02  & 2.40 $\pm$ 0.05 \\
RF Over.  & 350 & 93 & 0.83 $\pm$ 0.02 & 2.31 $\pm$ 0.02 & 2.34 $\pm$ 0.05\\
\end{tabularx}
\end{ruledtabular}
\end{table*}

A range of both coherence lengths and entanglement lengths were probed in these two experiments. As shown in Fig. \ref{fig:wavep}, as $\xi$ is reduced or $\Delta_t$ is increased, the neutron path states will increasingly overlap. As the beam coherence length must satisfy $\beta_t \leq \Delta_t $, we know that the following data were indeed obtained in the overlap regime: $\xi < \beta_t \leq \Delta_t$.

The contextuality witness was calculated using the protocol described in our previous paper \cite{shen2019}. In all cases, we found that the witness value was consistent with the maximum bound of the contextuality inequality times the measured neutron polarization Pol, namely $2\sqrt{2} \times$Pol, as shown in Table \ref{tab:WitValues}.
The distinguishable two-path assumption for the spatial subspace holds to a ratio of the entanglement length to the transverse beam coherence length of at least 0.24. 

\subsection{Contextuality inequality and its violation}
\label{CHSHwitness}
    
Our experiments prove the fundamental observation, behind the Kochen-Specker theorem \cite{ks-1967,mermin-rmp-1993}, that a quantum description of nature is necessarily contextual.  That is, measurement outcomes of compatible sets of quantum observables, known as contexts, cannot reveal pre-existing values of those properties measured. The measured values depend upon the context. 

In the present experiment, we use the CHSH inequality~\cite{chsh-1969} to test quantum contextuality of a neutron state in a particular experimental arrangement where the paths of the neutron are closer together than the transverse coherence length of the neutron beam. In Ref. \cite{lu2019operator} we developed the theory necessary to understand the way our neutron interferometers unveil the contextual nature of quantum reality by assuming a finite-dimensional Hilbert space representation, motivated at first by the conjecture that the neutron paths were non-overlapping. We have now established that contextuality applies even when the entanglement length is much less than the transverse coherence length of the neutron beam. Since we expect that the wavepacket size is \textit{larger} than the beam coherence length, we are obliged to extend our analysis to an infinite-dimensional Hilbert space. We associate to our system the tensor product Hilbert state space 
$\mathcal{H}=\mathcal{H}_{s} \otimes \mathcal{H}_{\br}$, 
where $\mathcal{H}_{s}$ describes a two-dimensional (spin-$\frac{1}{2}$) subspace while $\mathcal{H}_{\br}$ is the subspace spanned by the position of the neutron in $\mathds{R}^3$.

The most general state realized in RF flippers and MWPs entanglers,
\begin{equation}
    \label{Eqn:generalstate}
   \Phi_\text{n}(\br,t)=\frac{\phi_1(\br,t)\ket{\eta_1}+\phi_2(\br,t)\ket{\eta_2}}{\sqrt{2}},
\end{equation}
must be defined in $\mathcal{H}$. The entangled wave packet emerging after the entangler corresponds to a single neutron of momentum $\hat \bp$ (and mass $m_n$), characterized by the distribution $g(\bk)$ with mean wavevector $\bk_0$,  transverse spatial width $\Delta_t$, and energy $\bra{\Phi_\text{n}}\hat H_{\sf p}\ket{\Phi_\text{n}}=E_{\sf p}$, where $\hat H_{\sf p}=\frac{\hat \bp^2}{2 m_n}$,

\begin{eqnarray}
       \hspace{-0.5cm}  \Phi_\text{n}(\br,t)
    = \frac{1}{(2\pi)^{3/2}}\int d\bk \, g(\bk) e^{i \bk \cdot \br}e^{- i \omega(k) t} \ \ket{\eta_{\bk\cdot\bxi}(t)}, 
\end{eqnarray}
with dispersion $\omega(k)= \hbar k^2 / 2 m_n$ and generalized spin state

\begin{eqnarray}
\label{Eqn:wavefunction}
\ket{\eta_{\bk\cdot\bxi}(t)}&=&\frac{
    e^{- \frac{i}{2}\bk \cdot \bxi} \ket{\uparrow}+
    e^{\frac{i}{2}\bk \cdot \bxi}e^{-i \delta\omega(k) t} \ket{\downarrow}}{\sqrt{2}}.
\end{eqnarray}
The vector $\bxi$ denotes the separation between the paths with $|\bxi| = \xi$.
The two paths may be subject to different dispersion with $\delta\omega(k)$ being their difference, and the $z$-axis is the spin-quantization axis, so $\ket{\eta_1}=\ket{\uparrow}$ and $\ket{\eta_2}=\ket{\downarrow}$. We note that for the MWPs, $\delta \omega(k) \approx 0$ in Eqn. \eqref{Eqn:wavefunction}, as no quanta of energy is given to either branch to induce a spin flip ($\delta \omega(k) \neq 0$ as the guide field splits the Zeeman energies).

At $t=t_1$, i.e. after the neutron is exposed to spin- and path-phase shifters, its spin state evolves into 
\begin{equation}
    \ket{\eta_{\bk\cdot\bxi}(t_1)} = \frac{e^{- \frac{i}{2}\bk \cdot \bxi} \ket{\uparrow}
    + e^{\frac{i}{2}\bk \cdot \bxi} e^{i [\alpha+\chi-\delta\omega(k) t_1 + \theta_c(t_1)]} \ket{\downarrow}}{\sqrt{2}}.
\end{equation}
The additional phase $\theta_c(t)$ is introduced by the entangler and guide field and is removed by the disentangler when the neutron leaves the apparatus at time $t=t_2$. The spin state at the exit time $t_2$ is
\begin{eqnarray}
\label{Eqn:spinatt2}
    \ket{\eta_{\bk\cdot\bxi}(t_2)} &=& \frac{ \ket{\uparrow} + e^{i (\alpha+\chi)} \ket{\downarrow}}{\sqrt{2}} =
    \frac{ \ket{\uparrow} + e^{i \theta} \ket{\downarrow}}{\sqrt{2}},
\end{eqnarray}
with spin polarization ready to be detected. 

Finally, we measure the  neutron's spin polarization: $\bra{\Phi_{\sf n}(t_f)} \sigma^s \ket{\Phi_{\sf n}(t_f)}$ at $t=t_f$ when the wavepacket reaches the detector with spin state equal to Eqn. \eqref{Eqn:spinatt2}.
Experimentally, we only measure the $+x$-component of the spin, so the probability of recording a neutron count in the detector is
\begin{equation}
   \bra{\Phi_{\sf n}(t_f)} P_{+x} \ket{\Phi_{\sf n}(t_f)} = \frac{\cos(\theta)+1}{2} ,
\end{equation}
where $P_{+x} = \frac{(\ket{\uparrow} + \ket{\downarrow})(\bra{\uparrow} + \bra{\downarrow})}{2}$ is the usual $+x$-component spin projection operator. 

To connect this measurement to the witness $S$, defined below, we proceed as in Lu \cite{lu2019operator} and assume a distinguishable subsystem scenario for the path subspace. This distinguishability assumption of the path states is essential to derive the following CHSH contextuality inequality. 
The open question that this work addresses is whether this distinguishability requires $\xi>\Delta_t$.

We define two pairs of observables:  $\sigma^s_{u_i}$ and $\sigma^p_{v_j}$ acting on the spin and path subsystems, respectively, with $i,j\in\{1,2\}$, and $u(\alpha), v(\chi)$ labeling operators associated with angles $\alpha$ and $\chi$ in the $x$-$y$ plane of the corresponding Bloch spheres

\begin{subequations}
\begin{align}
    \label{Eqn:Def_Pauli_spin_path}
    \sigma_{u}^{s} &= \cos{\alpha}\, \sigma^{s}_x+\sin{\alpha}\, \sigma^s_y \\
    \sigma_{v}^{p} &=\cos{\chi}\, \sigma^{p}_x+\sin{\chi}\, \sigma^p_y,
\end{align}
\end{subequations}
where $\sigma^{s,p}_{x,y}$ are Pauli matrices. The distinguishable subsystem scenario is encapsulated in this choice of observables that permit the definition of the CHSH witness:
\begin{equation} \label{Eqn:CHSH Witness}
S=E(\alpha_1, \chi_1)+E(\alpha_1, \chi_2)+E(\alpha_2, \chi_1)-E(\alpha_2, \chi_2),
\end{equation}
where 
$E(\alpha, \chi)$ represent expectation values of $\sigma^s_u\sigma_v^p$ over a state $\ket{\Psi} \in {\cal H}$, i.e. 
\begin{equation}
E(\alpha, \chi)=E \left[ \sigma^s_{u(\alpha)}\sigma_{v(\chi)}^p \right]=
\bra{\Psi} \sigma^s_{u(\alpha)}\sigma_{v(\chi)}^p \ket{\Psi}.
\end{equation}

As shown in Lu \cite{lu2019operator}, the Pauli operators admit a projection operator decomposition, which also explains the structure of Eqn. \eqref{Eqn:expectationvalue}.
Because the phase shifters and entanglers can be represented by unitary operators, we can directly connect the experimental measurement to the expectation value defined above:
\begin{equation}\hspace*{-0.25cm}
    \bra{\Phi_{\sf n}(t_f)} P_{+x} \ket{\Phi_{\sf n}(t_f)}=2\bra{\Psi_{\sf Bell}} 
    P^s(\alpha) P^p(\chi) \ket{\Psi_{\sf Bell}}  ,
\end{equation}
where the projection operators are defined as
\begin{subequations}
\begin{align}
    P^s(\alpha) &= \frac{(\ket{\uparrow} + e^{i \alpha} \ket{\downarrow})(\bra{\uparrow} + e^{-i \alpha}\bra{\downarrow})}{2}, \\
    P^p(\chi) &= \frac{(\ket{\phi_1} + e^{i \chi}\ket{\phi_2})(\bra{\phi_1} + e^{-i \chi} \bra{\phi_2})}{2}
\end{align}
\end{subequations}
for particular phases $\alpha$ and $\chi$ chosen later in Eqn. \eqref{Eqn:expectationvalue}. The Bell state $\ket{\Psi_{\sf Bell}}=\ket{\Phi_{\sf n}(t<t_1)}$ corresponds to the state of the neutron immediately after it passes through the  entangler.

While arbitrary classical assignments of eigenvalues of observables by a local hidden-variable theory cannot violate the CHSH inequality
\begin{eqnarray} |S|\leq 2 ,
\label{CHSHineq}
\end{eqnarray}
quantum mechanical expectations can, with a maximum value for $S$ set by the Tsirelson bound
$2 \sqrt{2}$,
\begin{eqnarray}
    -2\leq &S& \leq 2 \ \ \ \ \ \ \ \ \ \   \mbox{(classical statistics)}   \nonumber\\
    -2\sqrt2\leq & S & \leq2\sqrt2  \ \ \ \ \ \  \mbox{(quantum statistics)}.    \nonumber
\end{eqnarray}
Any state violating the CHSH inequality \eqref{CHSHineq} is necessarily an entangled state in the spin and path degrees of freedom. We use the violation of such a test to prove that our neutron beam is entangled. Thus $S$ is the witness value, with any number larger than 2 proving the mode-entanglement of the neutron.

\subsection{Calculating Witness Values}

In order to measure the witness value $S$, the experimental data were used to extract the expectation values of Eqn. \eqref{Eqn:CHSH Witness}. The function $N_{\alpha,\chi}$ denotes the neutron counts in the detector for preset spin and path phases; $N_{\alpha,\chi}$ is normalized to overall beam intensity and corrected for quartz block transmission, and also corrected for background. The expectation values are calculated as in \cite{hasegawa2003,shen2019}:

\begin{equation}
    \label{Eqn:expectationvalue}
    E(\alpha,\chi)  = \frac{N_{\alpha,\chi} - N_{\alpha,\chi + \pi} - N_{\alpha + \pi,\chi} + N_{\alpha + \pi,\chi + \pi}}{N_{\alpha,\chi} + N_{\alpha,\chi + \pi} + N_{\alpha + \pi,\chi} + N_{\alpha + \pi,\chi + \pi}} .
\end{equation}

\begin{figure}[ht]
\includegraphics[width=3.4in]{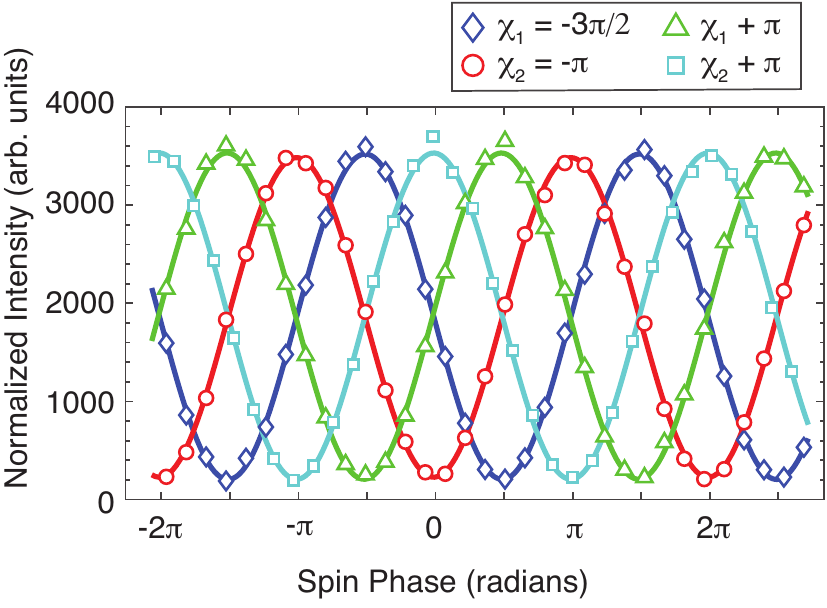} 
\caption{\label{fig:mwpanalysis} Transmission-corrected intensity vs. spin phase for four path phases in the 2 mm slit MWP experiment. Error bars corresponding to the statistical counting error are the size of the marker or smaller.}
\end{figure}

The maximum violation of the CHSH inequality is obtained when the spin and path angles satisfy $\alpha_1 + \chi_1 = -\pi/4$ and $\alpha_2-\alpha_1= \chi_2-\chi_1=\pi/2$. In the MWP experiment we chose values of $\alpha_1 = -3\pi/4$, $\alpha_2 = -\pi/4$, $\chi_1 = -3\pi/2$, and $\chi_2 = -\pi$. Neutron counts were measured at nine different values of $\chi$ equally spaced from $-\pi$ to $\pi$ (i.e. $-\pi$, $-3\pi/4$, $-\pi/2$, ..., $\pi$) and at around 30 equally spaced values of $\alpha$ (see Fig. \ref{fig:mwpanalysis}).
In the RF flipper experiment, the expectation values are obtained from the fitted intensity curves, so $\alpha_1$ may be chosen arbitrarily.
The function $N_{\alpha,\chi}$ is also fitted with a cosine wave from several path angles. 

In both experiments, imperfect neutron polarization reduces the value of $N_{\alpha, \chi}$ \cite{shen2019}. Therefore, the maximum observable value for the witness is expected to be $2\sqrt{2}\times$Pol; this value is also expected to be the value for a maximally entangled neutron. The polarization and both the experimentally determined and maximum witness values are shown in Table \ref{tab:WitValues}.

\subsubsection{MWP}
A single entanglement length was used for all the MWP measurements. The entanglement length and path phase calibration were determined from a fit of the path phases for various block angles using Eqn. \eqref{Eqn:pathphase}.  A witness value was found from scans of many spin phase angles at four different path phases, as shown in Fig. \ref{fig:mwpanalysis}. Each path phase was fitted with the following function:

\begin{equation}
\label{Eqn:N}
N_{\alpha,\chi} = C \cos(\alpha + \chi + \theta_0)+ D,
\end{equation}
where $C$ and $D$ are fitting parameters that include background and imperfect neutron polarization, and $\theta_0$ an additional constant phase originating from stray fields in the experimental setup. The intensity values used to calculate the expectation value $E(\alpha , \chi)$ in Eqn. \eqref{Eqn:expectationvalue} were evaluated using values obtained from these fitting functions. A witness value of 2.50$\pm$0.01 was found for $\beta_t$ values of 550 nm, 140 nm, and 70 nm, consistent with $2\sqrt{2} \times$Pol. A Monte Carlo simulation was used to calculate the statistical uncertainty \cite{shen2019}. As an aside, the witness value is independent of the $\chi$-dependent transmission value provided that Eqn. \ref{Eqn:N} holds and the polarization remains constant (see the Appendix for more details). With these assumptions, the non-transmission-corrected data would yield the same witness value.

\subsubsection{RF flippers}
On the time of flight RF flipper experiment, for each combination of spin and path phase, the intensity was measured at the detector and normalized to the quartz block transmission. 
The spin and path phases were calibrated by fitting the wavelength dependence of the time-of-flight polarization data \cite{shen2019}. The neutron polarization was measured as a function of wavelength and normalized to the polarization when no phase is applied ($\alpha=\chi=0$). The normalized polarization is well fitted over the wavelength range 3.8 to 8.0 {\AA} by the equation
\begin{equation}
\mathrm{Pol} = \frac{\cos[(\alpha-\alpha_0)\lambda_n + b\lambda_n^3+\varphi_{\rm RF}]}{\cos(\alpha_0\lambda_n-\varphi_{\rm RF})}  ,
\end{equation}
where $\alpha_0$ and $\varphi_{\rm RF}$ account for small tuning errors of the echo condition and of the RF flipper phases, respectively \cite{shen2019}. The relative path phase was fitted by its $\lambda_n^3$ dependence and found to match the phase calculated from the instrument parameters to within 3.6$^\circ$.
The intensities are fitted with Eqn. \eqref{Eqn:N} as well.
The statistical errors quoted for the witnesses are standard deviations arising only from the propagation of counting statistics. A slight difference in polarization between the initial spin up and spin down states may lead to a slightly different witness value between these measurements. In contrast to our earlier experiment, we did not observe any evidence for systematic errors. Witness values of 2.42 $\pm$ 0.02 and 2.35 $\pm$ 0.02 were found for the RF conventional mode and RF overlap mode, respectively, consistent with $2 \sqrt{2} \times$Pol.

\FloatBarrier
\section{Discussion}

There are two primary takeaways from these experiments: neither varying the entanglement length $\xi$ nor the velocity difference between the wavepacket branches reduced the entanglement witness value, and the neutron's spin and path subsystems were entangled using both MWPs and RF flippers in multiple configurations. Moreover, the paths can be treated as distinguishable even when the entanglement length is less than the transverse beam coherence length $\beta_t$. By decreasing the ratio ${\xi}/{\beta_t}$ to much less than unity, we have shown that maximal entanglement persists past the point where the path branches certainly overlap. The introduction of overlap mode for the RF flippers also showed that the contextuality witness does not depend on whether the branches of the wavepacket significantly overlap one another as they pass through the path-phase crystal. 

The modular nature of these entangler-disentangler pairs was crucial in proving the quantum contextual nature of our carefully prepared neutron beam through the construction of a specifically chosen entanglement witness $S$. These two results combine to show the robustness of the neutron subsystems entanglement and its potential suitability as a universal probe of quantum materials. 

The applications of an entangled neutron probe may depend on what entanglement lengths are achievable. The range of entanglement lengths $\xi$ available depends solely on the instrument; to our current knowledge, $\xi$ can range from over 20 microns down to a few tens of nanometers, with the upper bound set by the maximum field strength in the entangler-disentangler pair and the lower bound set by instrumental aberrations.

\begin{figure}[t]
\includegraphics[width=3.4in]{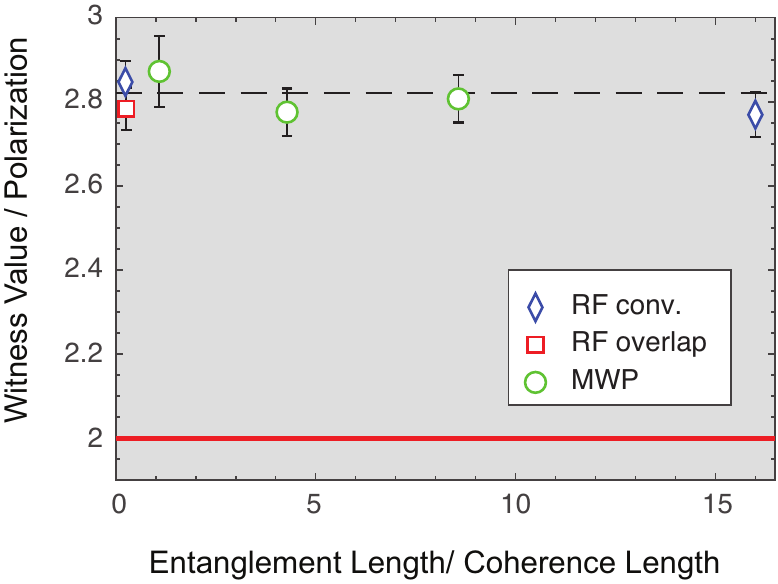} 
\caption{\label{fig:witnessvalues} The witness value divided by the polarization ($S/{\rm Pol}$) vs. entanglement length divided by coherence length ($\xi/\beta_t$). Data is taken from Tab. \ref{tab:WitValues}. The dashed line at the Tsirelson bound 2$\sqrt{2}$ is the expected witness value for a maximally entangled neutron. The red line at 2 is the upper bound for any classical, non-contextual, theory. }
\end{figure}

As to the second takeaway, we note that the MWP and the RF flipper both split the path states by refraction but that the RF flipper also causes a change in the neutron’s total energy. Despite the different underlying mechanisms, both devices maximally mode-entangle the neutron subsystems leading to identical results. The equivalent witness values $S$ for the two experiments also emphasize that our results are independent of the neutron beam preparation, since we used both a reactor source and a pulsed spallation source. We note that the MWP and RF flippers are both compatible with reactor and spallation sources. Thus, we expect that adding either MWPs or RF flippers to any existing polarized neutron beamline could generate an entangled neutron probe; this ability shows the great flexibility and universality of our entangling devices.

We have recently developed a general quantum entangled-probe scattering theory \cite{Irfan2020}, which establishes the framework to respond to the obvious question: what kind of information can be extracted with this novel probe?
Interestingly, by carefully tuning the probe's entanglement and intrinsic coherence properties, one can directly view the inherent entanglement of the target material. This theoretical framework supports the view that our entangled beam could be used as a multipurpose scientific tool. In this regard, we are currently considering several ideas for future experiments which could include measuring the physical sizes of Cooper pairs in different superconductors and imaging edge states in topological insulators. Samples that display long-range 1D magnetic order on the tens of nm scale, such as Heisenberg spin chains, are especially interesting.  

Mode-entanglement has also been produced and controlled in other experiments.
Entanglement between the polarization and path of a single photon in a traditional Mach-Zehnder interferometer was quantified using the CHSH and Clauser and Horne (CH) inequalities, although the results were not consistent with maximal entanglement between these degrees of freedom \cite{Bellphoton2009}.
Mode-entanglement between the hyperfine spin state and motional state of trapped $^{171}$Y$^+$ ions has also been demonstrated by measuring the contrast of Ramsey fringes, although there was no quantitative determination of the degree of entanglement \cite{singleatomint-2013}.
Traditional neutron interferometers have also entangled the spin, path, and energy of a single neutron, and have also quantified the degree of entanglement via the violation of the CHSH inequality \cite{fundQM-2014}.

We now contrast this technique to similar work with entangled photons. Our neutron beam is mode-entangled and not particle-entangled \cite{barnum-2004}. Indeed, any entanglement study is naturally compared to measurements on particle entanglement as is in the case of a beam of photon pairs. A key difference is that here a single particle (the neutron) has entangled subsystems, instead of the entanglement occurring in a spatially separated pair. One could in principle produce entangled photon beams which are sensitive to multi-point correlation functions \cite{Schotland2016,Irfan2020}. However, due to the stochastic nature of production of neutrons, this type of entanglement seems unlikely.

Our results regarding the insensitivity of the entanglement witness value that we measure to our assumptions regarding the neutron wave packets is consistent with other previous results in the literature. A very simple argument by Stodolsky~\cite{Stodolsky1998} shows that the detector intensity measurements of the type our entanglement witness is ultimately constructed from do not depend on assumptions about whether or not the incident beam is composed of coherent wave packets. If the source of particles is stationary then the only thing the density matrix of the incoming beam can depend on is the energy/momentum spectrum. Although technically speaking the source of neutrons in this experiment is pulsed and therefore time-dependent, the timescale is so slow compared to any possible wave packet dynamics that the source can be taken to be quasi-stationary in our experiment, with the pulsed nature of the neutron source used only to determine the mean neutron speed in the beam at any instant. 

In a series of atom interferometry experiments by the Pritchard group at MIT they constructed a \lq\lq detuned separated oscillatory field'' longitudinal atom interferometer (in the neutron scattering/optics world this device would call a MIEZE spectrometer). They showed experimentally the absence of coherent wave packets in their source, which similar to our source of neutrons comes ultimately from a nearly-thermalized ensemble, by looking for off-diagonal density matrix components in the atom beam that fed the interferometer~\cite{Rubenstein1999a}. They also measured the presence of off-diagonal components of the density matrix of the beam upon introducing time-dependent modulations in the source upstream of the interferometer~\cite{Rubenstein1999b} on a sufficiently fast timescale to violate Stodolsky's assumptions.  These results were all consistent with their quantum mechanical treatment of their interferometer~\cite{Pritchard1999}. Other examples of theoretical treatments of systems which introduce nonstationary elements at or downstream of the source and therefore can in principle say something about the wave packet structure of the incident beam are the work of Golub and Lamoreaux~\cite{Golub1992},  where they point out a way to  measure the transverse components of a neutron wave packet, and the work of  Robicheaux and Noordam~\cite{Robicheaux2000} on pulsed electron scattering.

For additional perspective on the meaning of our result on the preservation of the entanglement witness values under the different experimental conditions considered in this work, it is illuminating to look at atom interferometry experiments~\cite{Cronin2009}, which tested various types of decohering interactions in the interferometer. Given the higher sensitivity of atoms to environmental perturbations compared to neutrons due to their stronger coupling to the electromagnetic field, it is easier to investigate such questions experimentally with atoms. Consistent with the laws of quantum mechanics, experiments found that as long as nothing about the apparatus and/or the environment is ``labeling'' any of the paths in the relevant qubit subspaces, which is the case for the coherent interactions with matter and external fields that the neutrons in our experiment were subjected to, then the interferometer contrast, and therefore the entanglement witness values that are constructed from the amplitudes after the different phase shifts are applied to the different qubit subspaces, is unchanged.

\FloatBarrier
\section{Conclusion}

We have developed a quantum probe consisting of single particle mode-entangled neutrons. In these experiments, the individual neutrons' spin and path distinguishable subsystems were entangled. The observed violation of the CHSH contextuality inequality proved that the beam was maximally entangled in both experiments. There was no difference in the degree of entanglement between using MWPs or RF flippers or performing the experiment at a continuous reactor source or a pulsed spallation source.

Furthermore, the separation of the path states can be reduced to tens of nanometers without imposing a similarly restrictive constraint on the beam coherence length; access to this length scale is required for probing correlations in many complex quantum materials. Additionally, neither an energy difference nor a longitudinal spatial separation at the sample position will degrade the probe. From these results, we conclude that we now have access to a robust, tunable entangled probe, suitable for exploring exotic excitations in entangled matter.

\FloatBarrier
\section{Acknowledgements}
We thank Lee Robertson and Lowell Crow for their help with the HFIR experimental setup. We also appreciate the useful discussions with J. Fernandez-Baca. We thank Jeroen Plomp and Michel Thijs for assistance setting up the RF coils and guide fields and useful discussions. 

A portion of this research used resources at the High Flux Isotope Reactor, a DOE Office of Science User Facility operated by the Oak Ridge National Laboratory. This work is sponsored by the Laboratory Directed Research and Development Program of Oak Ridge National Laboratory, managed by UT-Battelle, LLC, for the U. S. Department of Energy. This material is based upon work supported by the U.S. Department of Energy, Office of Science, Office of Basic Energy Sciences under contract number DE-AC05-00OR22725. The work described in this paper arose from the development of magnetic Wollaston prisms funded by the US Department of Energy through its STTR program (grant number DE-SC0009584). A number of the authors acknowledge support from the US Department of Commerce through cooperative agreement number 70NANB15H259. 

Experiments at the ISIS Neutron and Muon Source were supported by a beamtime allocation RB1920268 from the Science and Technology Facilities Council (DOI: 10.5286/ISIS.E.RB1920268).
N. Geerits acknowledges a grant from the Austrian Science Fund (FWF) Project No. P30677 and P34239.
W. M. Snow acknowledges support from US National Science Foundation grant PHY-1914405 and the Indiana University Center for Spacetime Symmetries. 
The IU Quantum Science and Engineering Center is supported by the Office of the IU Bloomington Vice Provost for Research through its Emerging Areas of Research program.

\FloatBarrier
\section{Appendix}

Table \ref{tab:LongWavelengthWitness} lists witness values at various neutron wavelengths; these witnesses were obtained with RF flippers using the Larmor instrument.

\begin{table}[h]
\centering
\caption{\label{tab:LongWavelengthWitness} ISIS witness values and polarizations at various wavelengths for both the conventional (conv.) and overlap mode.}
\begin{ruledtabular}
\newcolumntype{R}{>{\centering\arraybackslash}X}
\begin{tabularx}{0.181\linewidth}{R|R|R|R|R}
   Wave-length (\AA) & Conv. Mode Witness Value & Conv. Mode Pol. & Overlap Mode Witness Value & Overlap Mode Pol. \\
\hline
\rule{0pt}{2.5ex} 
4 & 2.42 $\pm$ 0.02 & 0.85  $\pm$ 0.02 &2.31 $\pm$ 0.02 & 0.83 $\pm$ 0.02\\
\hline
\rule{0pt}{2.5ex} 
4.5 & 2.39 $\pm$ 0.02 & 0.84 $\pm$ 0.02 &2.29 $\pm$ 0.02 & 0.79 $\pm$ 0.02\\
\hline
\rule{0pt}{2.5ex} 
5 & 2.35 $\pm$ 0.02 & 0.82 $\pm$ 0.03 &2.26 $\pm$ 0.02 & 0.80 $\pm$ 0.02\\
\hline
\rule{0pt}{2.5ex} 
5.5 & 2.38 $\pm$ 0.02& 0.84 $\pm$ 0.03 &2.29 $\pm$ 0.02 & 0.82 $\pm$ 0.03\\
\hline
\rule{0pt}{2.5ex} 
6 & 2.32 $\pm$ 0.02& 0.79 $\pm$ 0.04 &2.26 $\pm$ 0.03 & 0.80 $\pm$ 0.03\\
\hline
\rule{0pt}{2.5ex} 
6.5 & 2.25 $\pm$ 0.03& 0.78 $\pm$ 0.05 &2.23 $\pm$ 0.04 & 0.79 $\pm$ 0.04\\
\hline
\rule{0pt}{2.5ex} 
7 & 2.22 $\pm$ 0.04& 0.78 $\pm$ 0.05 &2.21 $\pm$ 0.04 & 0.78 $\pm$ 0.05\\
\end{tabularx}
\end{ruledtabular}
\end{table}

We now discuss the effect of transmission on the calculated witness value in more detail. First, we assume the following: (a) intensity depends only on the cosine of the sum of $\alpha$, $\chi$, and some constant phase $\theta_0$, (b) background, polarization, and incident flux are constant, (c) transmission is only $\chi$ dependent, and (d) nearly all of the beam passes through both quartz blocks. With these assumptions, let $N'_{\alpha, \chi}$ be the non-transmission-corrected neutron counts recorded at the detector, so 
\begin{align}
    \label{Eqn:N'}
    N'_{\alpha,\chi} &= \frac{1}{2} I_0 \mathrm{T}(|\chi|) [1 + \mathrm{Pol} \times \cos(\alpha + \chi + \theta_0)] + \mathrm{BG}
\end{align}
where $I_0$ is the incident flux, $\mathrm{T}(|\chi|)$ is the transmission for a particular path phase, $\mathrm{Pol}$ the beam polarization, and $\mathrm{BG}$ is the background. We note that $0 \leq \mathrm{T}(|\chi|) \leq 1$ and the transmission is an odd function of $\chi$ due to our experimental setup (see Fig. \ref{fig:instrument}). Once transmission corrected, Eqn. \eqref{Eqn:N'} is equivalent to Eqn. \eqref{Eqn:N} with suitable choices of the $C$ and $D$ coefficients. Using this model, we find that each expectation value defined in Eqn. \eqref{Eqn:expectationvalue} are transmission independent if $I_0 \gg B$, which is the case for our experiments.

\begin{figure}[ht]
\includegraphics[width=3.4in]{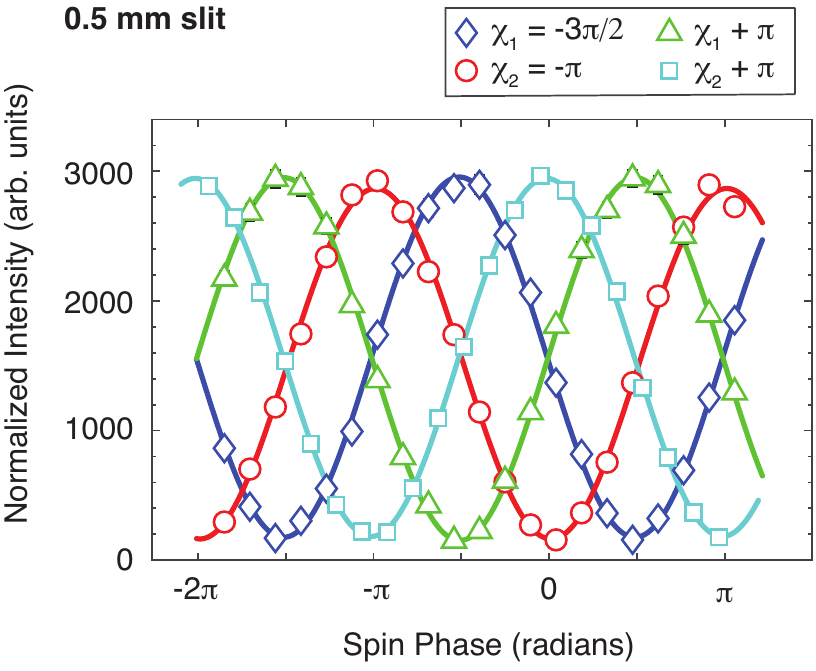} 
\caption{\label{fig:HFIRappendixdata} Transmission-corrected intensity vs. spin phase for four path phases in the 0.5 mm slit MWP experiment. Error bars corresponding to the statistical counting error are the size of the marker or smaller except where shown.}
\end{figure}

\begin{figure*}[ht]
\includegraphics[width=7in]{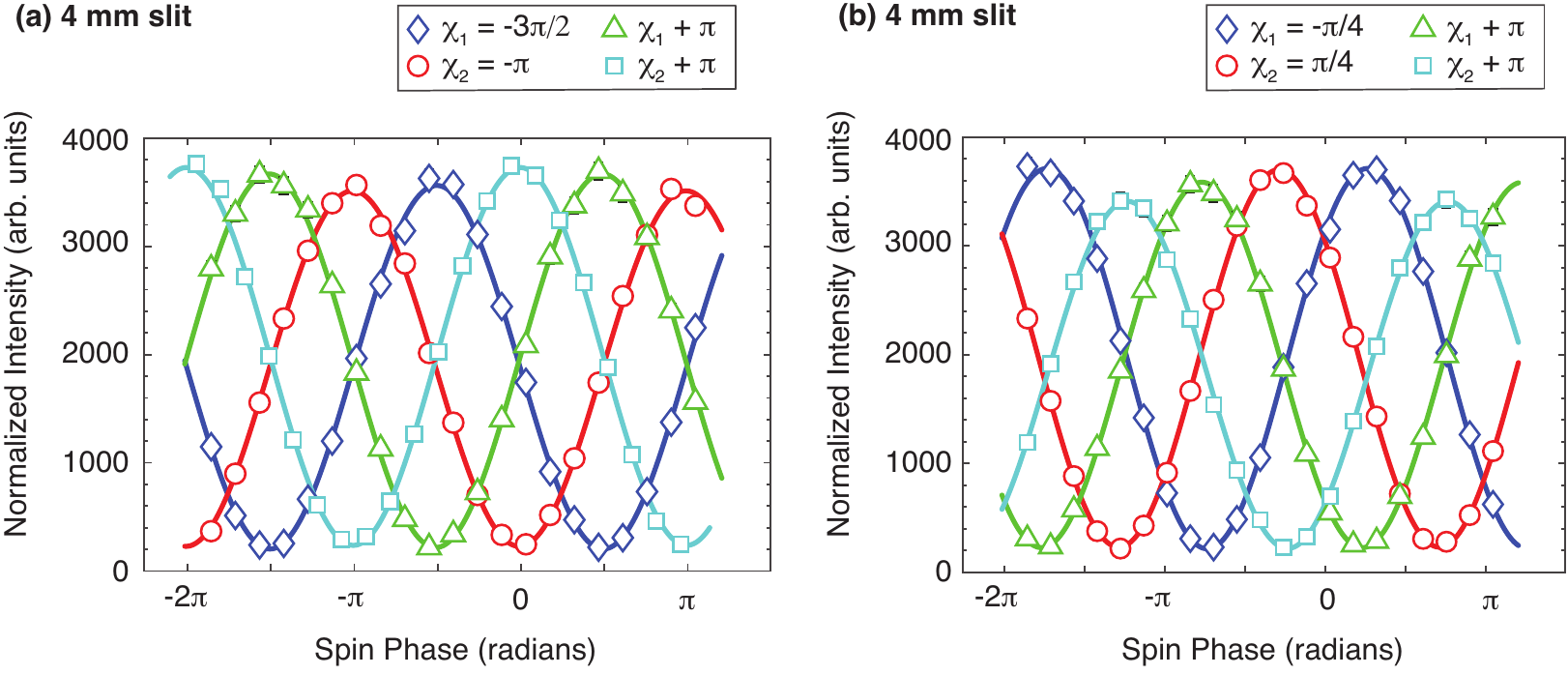} 
\caption{\label{fig:HFIRappendixdata2} Transmission-corrected intensity vs. spin phase for four path phases in the 4 mm slit MWP experiment. (a) and (b) show two different $\chi_1$ values. Error bars corresponding to the statistical counting error are the size of the marker or smaller except where shown.}
\end{figure*}

Figures \ref{fig:HFIRappendixdata} and \ref{fig:HFIRappendixdata2} show the fit of the intensities for the 0.5 mm and 4 mm slit widths, respectively. For the 4 mm data, the polarizations of the $\chi$ values are consistent within error. We attribute the variation in the peak intensities in Fig. \ref{fig:HFIRappendixdata2} to a small part of the beam missing the second quartz block. From the argument in the previous paragraph, the witness value is still consistent with maximal entanglement.

\bibliographystyle{apsrev4-2.bst}
\bibliography{sources.bib}
\end{document}